\documentclass[aps,prl,twocolumn,showpacs,floatfix,superscriptaddress,amsmath,amssymb]{revtex4-2}

\usepackage{amsfonts}
\usepackage{mathrsfs}
\usepackage{amsmath}
\usepackage{xcolor}
\usepackage{graphicx}
\usepackage{bm}
\usepackage{amssymb}
\usepackage{physics}
\usepackage{xspace}
\usepackage{epstopdf}
\usepackage{dcolumn}
\usepackage{longtable}
\usepackage{multirow}
\usepackage{float}
\usepackage{array}
\usepackage{comment}
\usepackage[colorlinks=true, letterpaper=true, pdfstartview=FitV,  linkcolor=blue, citecolor=blue, urlcolor=blue]{hyperref}

\makeatletter

\providecommand{\tabularnewline}{\\}

\makeatother

\begin{document}
\title{Eccentricity valley Hall effect}

\author{Jin Cao}
\affiliation{Research Laboratory for Quantum Materials, Department of Applied Physics, The Hong Kong Polytechnic University, Kowloon, Hong Kong, China}

\author{Shen Lai}
\affiliation{Institute of Applied Physics and Materials Engineering, Faculty of Science and Technology, University of Macau, Macau, China}

\author{Cong Xiao}
\email{congxiao@fudan.edu.cn}
\affiliation{Interdisciplinary Center for Theoretical Physics and Information Sciences (ICTPIS), Fudan University, Shanghai 200433, China}

\author{Qian Niu}
\affiliation{Department of Physics, University of Science and Technology of China, Hefei, Anhui 230026, China}

\author{Shengyuan A. Yang}
\email{shengyuan.yang@polyu.edu.hk}
\affiliation{Research Laboratory for Quantum Materials, Department of Applied Physics, The Hong Kong Polytechnic University, Kowloon, Hong Kong, China}

\begin{abstract}
Valleytronics harnesses the valley degree of freedom --- energy-degenerate extrema in the electronic band structure --- for information storage and processing. Valley Hall effect (VHE) is a cornerstone of valleytronics, enabling electric generation of pure valley currents.
While extensively studied in systems with valleys located at time-reversal-breaking points,
here, we shift the paradigm to valleytronic platforms with time-reversal-invariant valleys (TRIVs), revealing a novel phenomenon: eccentricity VHE.
Unlike conventional VHE, the valley Hall angle for eccentricity VHE is an intrinsic geometric property, governed solely by the eccentricity of the valley Fermi surface, rendering it highly robust against variations in temperature or carrier density.
Eccentricity VHE emerges universally across all 25 layer groups supporting TRIVs. We demonstrate these distinctive features in monolayer GeS$_{2}$ via first-principles calculations, predicting a significant valley Hall angle of 0.74.
This effect can be detected through nonlocal transport measurements exhibiting characteristic scaling behavior, or, in certain cases, through valley-layer coupling. Our findings reveal a critical overlooked facet of valley Hall physics, transcend the established VHE paradigm, and significantly broadens the scope of valleytronics.
\end{abstract}

\maketitle

\begin{figure*}
\begin{centering}
\includegraphics[width=14cm]{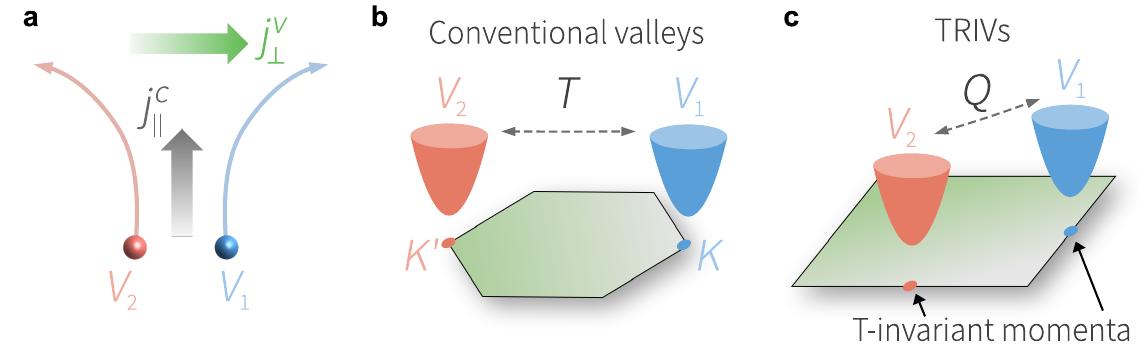}
\par\end{centering}
\caption{\label{fig1}\textbf{Valley Hall effect and two categories of valleytronic systems.} \textbf{a},
In VHE, a longitudinal charge current $j^c_\|$ drives a transverse valley current $j^v_\bot$. The blue and red spheres denote electrons from two different valleys $V_1$ and $V_2$. \textbf{b}, Conventional valleytronic systems, such as graphene and transition metal dichalcognides materials, have a pair of valleys located at $K$ and $K'$ points of the hexagonal Brillouin zone. The two valleys are related by time-reversal $\mathcal T$, yet each valley is at a $k$ point that does not preserve $\mathcal T$. \textbf{c}, Valleytronic system with time-reversal-invariant valleys (TRIVs). Each valley here resides at a $\mathcal T$-invariant momentum point. The two valleys are not related by $\mathcal T$ but by some crystalline symmetry $\mathcal{Q}$.}
\end{figure*}

In the valley Hall effect (VHE), a longitudinal charge current $j^c_\|$ induces a valley current $j^v_\bot$ flowing in the transverse direction, as illustrated in Fig.\,\ref{fig1}a, providing a key mechanism for purely electrical charge-valley conversion~\cite{Xiao2007Valley,Xiao2012Coupled,Mak2014valley,Gorbachev2014Detecting,Sui2015Gate,Shimazaki2015Generation,Lee2016Electrical,Wu2019Intrinsic}.
To date, the research on VHE, as well as the whole valleytronics field, has primarily focused on two-dimensional (2D) hexagonal-lattice materials, with graphene and transition metal dichalcogenides serving as paradigmatic examples~\cite{Rycerz2007Valley,Xiao2007Valley,Yao2008Valley,Xiao2012Coupled,Cai2013Magnetic,Mak2014valley,Gorbachev2014Detecting,Sui2015Gate,Shimazaki2015Generation,Lee2016Electrical,Wu2019Intrinsic,hung2019direct,Wang2020room,jiang2022room,Xu2014Spin,Lee2016Electrical,Schaibley2016Valleytronics,Vitale2018Valleytronics,Mak2018Light}. In these conventional valleytronic systems, the two valleys reside at the corners $K$ and $K'$ of the hexagonal Brillouin zone.
They are related by the time-reversal symmetry $\mathcal T$, yet each lies at
a momentum point that does not preserve $\mathcal T$ (see Fig.\,\ref{fig1}b). The VHE in such systems requires breaking inversion symmetry $\mathcal P$, and
it acquires a Berry-phase contribution~\cite{Xiao2007Valley,Xiao2012Coupled}.
The associated valley Hall angle $\theta_\text{VH} \equiv |j^v_\bot/j^c_\| |$, a key indicator of VHE efficiency, scales as $\tau^{-1}$ with the scattering time $\tau$, and  depends sensitively on parameters such as temperature and carrier density~\cite{Gorbachev2014Detecting,Sui2015Gate,Shimazaki2015Generation}.

Here, we propose a paradigm shift to a novel class of valleytronic platforms featuring time-reversal-invariant valleys (TRIVs), where the valleys reside  at $\mathcal T$-invariant momentum points (see Fig.\,\ref{fig1}c). In contrast to conventional valleytronic systems,
each TRIV is preserved under $\mathcal T$ (and $\mathcal P$) operation, leading to profound differences in valley properties. Crucially, valley transport is no longer prohibited by $\mathcal P$ symmetry. More importantly,  symmetry dictates that it arises from fundamentally distinct microscopic mechanisms. We predict a novel eccentricity VHE, in which the valley Hall angle $\theta_\text{VH}$ is governed solely by an \emph{intrinsic geometric} parameter --- the eccentricity $\mathfrak{e}$ of valley Fermi surface. As a result, $\theta_\text{VH}$ becomes independent of scattering time $\tau$ and exhibits exceptional robustness against variations in temperature and carrier density. Symmetry analysis further reveals that eccentricity VHE is universal across all 25 layer groups compatible with TRIVs. These distinctive features are explicitly demonstrated in monolayer GeS$_2$, where we predict a giant $\theta_\text{VH}\sim 0.74$.
Experimental signatures of eccentricity VHE can be identified through its unique scaling behavior in nonlocal transport or via gate-field control through valley-layer coupling.
Our proposal uncovers a previously unexplored frontier in valleytronics and may open the door to extensive research in related directions.

\paragraph{\textcolor{blue}{Symmetry character of VHE.}}
Let us consider a 2D nonmagnetic system having
two and only two energy-degenerate valleys, labeled as $V_1$ and $V_2$.
Such systems can be fully classified into two categories: (i) those with $\mathcal{T}$-connected valleys and (ii) those with TRIVs. Category (i) just corresponds to the conventional case.

Consider the valley current driven by applied $E$ field:
$
  j_{a}^{v}\equiv j_a^{V_1}-j_a^{V_2}=\sum_b\sigma_{ab}^{v}E_{b},
$
where roman subscripts denote Cartesian components, $j^{V_i}$ is the current from carriers in valley $V_i$,
and the valley conductivity
\begin{equation}
  \sigma^{v}=\sigma^{V_{1}}-\sigma^{V_{2}},
\end{equation}
with $\sigma^{V_i}$ being the valley-resolved charge conductivity.
Since the system respects $\mathcal T$ symmetry, the overall response coefficients must
be even under $\mathcal T$ operation, e.g., $\sigma^v$ is a $\mathcal T$-even quantity. Nevertheless, the
valley-resolved quantities like $\sigma^{V_i}$ is not constrained to have a definite parity under $\mathcal T$; in general, it contains both $\mathcal T$-even and $\mathcal T$-odd components, which have the following transformation behavior under $\mathcal T$:
\begin{equation}\label{Tcond}
  \mathcal T:\ \sigma^{V_i}_\eta\rightleftharpoons \eta \sigma^{\mathcal T V_i}_\eta,
\end{equation}
where $\eta=\pm $  for $\mathcal T$-even/odd component.

Now, one can see why the two categories make a salient difference here. For the conventional category (i) systems,
$\mathcal T$ switches the two valleys, i.e., $\mathcal T V_1=V_2$. Since $\sigma^v$ is odd under switch of valley indices, one finds only the $\mathcal T$-odd component $\sigma^{V_i}_-$ of $\sigma^{V_i}$ contributes to $\sigma^v$, and we have $\sigma^v=2\sigma^{V_1}_-$.

In contrast, for category (ii) with TRIVs, $\mathcal T$ operation preserves each valley, i.e.,
$\mathcal T V_i=V_i$. It follows from (\ref{Tcond}) that $\sigma^{V_i}$ contains only $\mathcal T$-even component,
so only
$\mathcal T$-even contributions enter the valley conductivity $\sigma^v$, and we may write $\sigma^v=\sigma^{V_1}_+ -\sigma^{V_2}_+$.

The revealed distinct symmetry characters indicate valley transport in the two categories must involve distinct mechanisms.
For example, it is well known that the conventional VHE, which occurs in category (i) systems, contains a Berry-phase contribution~\cite{Xiao2007Valley}, which is indeed $\mathcal T$-odd in each valley. However, it cannot contribute to VHE in category (ii) TRIV systems. Instead, we shall show below that valley transport in TRIV systems is mainly from the Drude contribution.

In addition, the two categories also exhibit distinct behaviors in the presence of inversion symmetry.
For category (i), the constraint of $\mathcal P$ requires $\sigma^{V_1}=\sigma^{V_2}$, so a nonzero valley transport must require the breaking of $\mathcal P$ symmetry, as is well-known for conventional VHE. In comparison, $\mathcal P$ has no constraint on $\sigma^v$ in category (ii), indicating that VHE can in principle be realized in TRIV systems with preserved $\mathcal P$ symmetry. This point will be confirmed in a while.

\paragraph{\textcolor{blue}{Eccentricity VHE.}}
For TRIV systems, we predict a new type of VHE. Remarkably,
not only the response structure but also the expression of this VHE have a rather generic form, leading to a purely \emph{geometric} valley Hall angle.

\begin{figure}
\begin{centering}
\includegraphics[width=8.7cm]{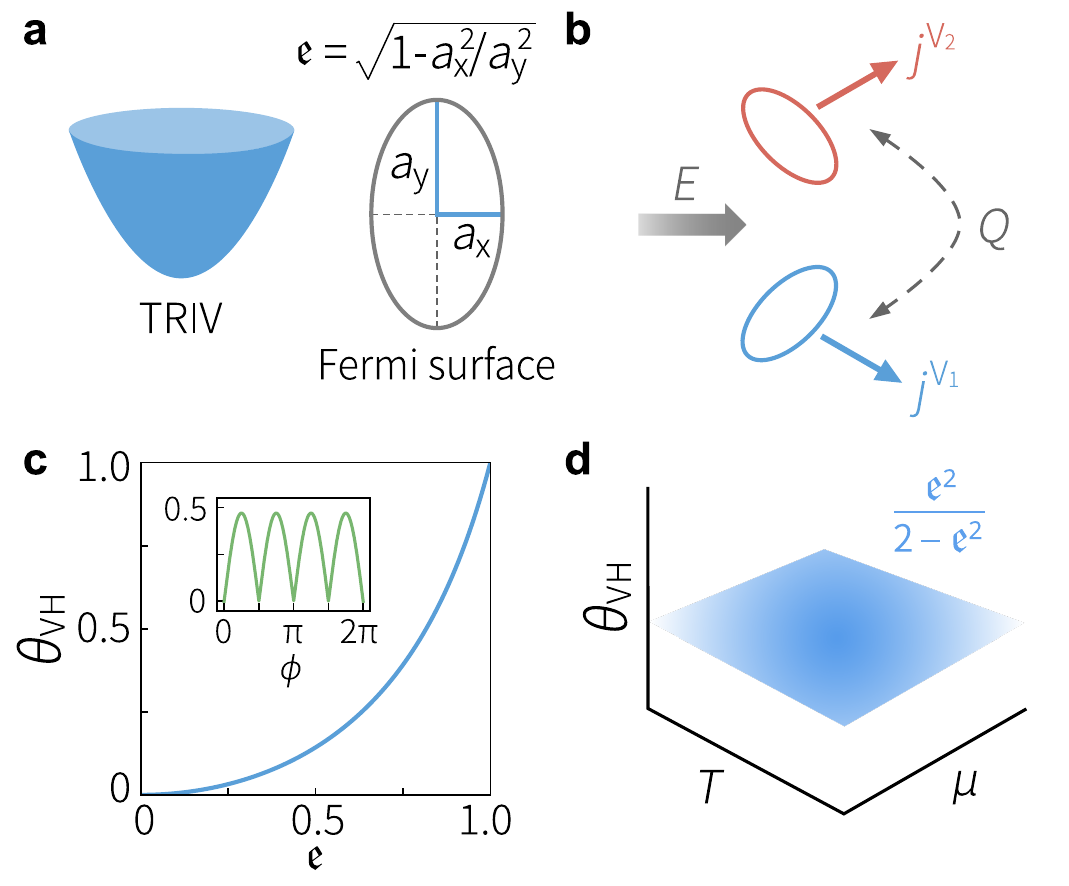}
\par\end{centering}
\caption{\label{fig2}\textbf{Eccentricity valley Hall effect.} \textbf{a}, The generic shape of a TRIV Fermi surface is an ellipse. Its geometry is characterized by the eccentricity $\mathfrak{e}$. Here, we illustrate the case of a TRIV Fermi surface with semi-major axis along $y$. \textbf{b}, Illustration of a pair of TRIVs connected by $\mathcal Q=C_{4z}$ symmetry. Under a driving electric field, the valley-contrasted current response leads to a VHE determined by eccentricity of valley Fermi ellipse. \textbf{c}, The resulting valley Hall angle as a function of $\mathfrak{e}$ and (inset) of the orientation of the applied electric field. \textbf{d}, The eccentricity VHE features a valley Hall angle independent of temperature and chemical potential.}
\end{figure}

Without loss of generality, let us consider TRIVs at conduction band bottom of a semiconductor band structure.
Constrained by $\mathcal T$ symmetry, the effective Hamiltonian $\mathcal H^{V_1}$ expanded around the center of valley  $V_1$
is generally of a quadratic form in the momenta $\bm k=(k_x, k_y)$. By choosing proper coordinate axis, one can always diagonalize this form, after which
$\mathcal H^{V_1}$ takes the following generic form
\begin{equation}
\mathcal H^{V_1}=k_{x}^{2}/a_x^{2}+k_{y}^{2}/a_y^{2}-\mu,\label{ham}
\end{equation}
where parameters $a_x$ and $a_y$ manifest the effective mass along the two principal axis, and $\mu>0$ is the chemical potential. The generic shape of Fermi surface at such a TRIV is an ellipse, characterized by the ratio $\lambda$ between the semi-major and semi-minor axis. Figure~\ref{fig2}(a) shows the case when the semi-major axis is along $y$, for which $\lambda=a_y/a_x$.

The $\mathcal T$-even $\sigma_+^{V_1}$, which is the component that can contribute to valley transport for TRIVs, is dominated by the Drude contribution. It is expressed as (we drop the subscript $+$ below and take $e=\hbar=1$) $\sigma_{ab}^{V_i}=-\tau \int_{V_i} v_{a}\partial_{b}f_{0}$, where $v_a$ is the band velocity, and $f_0$ is the Fermi distribution. For $V_1$ described by Eq.~(\ref{ham}), one easily obtains
$
\left(\sigma_{xx}^{V_1},\sigma_{yy}^{V_1}\right) = \frac{1}{2\pi}\mu\tau(\lambda,\lambda^{-1}).
$

To have a well-defined binary valley degree of freedom, the other valley $V_2$ must be connected to $V_1$ by some crystalline symmetry $\mathcal Q$ (see Fig.~\ref{fig1}c). For instance, assume the two valleys are connected by $\mathcal Q=C_{4z}$ symmetry (Fig.~\ref{fig2}b), then we have
$\sigma_{xx}^{V_2}=\sigma_{yy}^{V_1}$ and $\sigma_{yy}^{V_2}=\sigma_{xx}^{V_1}$. Under an in-plane electric field $\boldsymbol{E}=E\left(\cos\phi,\sin\phi\right)$, the induced charge current is purely longitudinal, i.e., parallel with $E$ field, given by
\begin{eqnarray}
j_{\|}^c & = & \frac{1}{2\pi}\mu\tau \left(\lambda+\lambda^{-1}\right)E.
\end{eqnarray}
Meanwhile, in the transverse direction, the induced VHE current  is
\begin{eqnarray}
j_{\bot}^{v} & = & \frac{1}{2\pi}\mu\tau \sin\left(2\phi\right)\left(\lambda-\lambda^{-1}\right)E,
\label{VHC}
\end{eqnarray}
which exhibits a $\pi$-periodic angular dependence on the field direction.

Remarkably, the valley Hall angle in this case has an especially simple form and can be expressed in terms of the eccentricity $\mathfrak e\ (=\sqrt{1-\lambda^{-2}})$ of the valley Fermi surface:
\begin{eqnarray}
\theta_{\mathrm{VH}}=\frac{\mathfrak e^2}{2-\mathfrak e^2}\left|\sin\left(2\phi\right)\right|.
\label{VHA}
\end{eqnarray}
This result is purely geometric: For a given driving field direction $\phi$, $\theta_{\mathrm{VH}}$ is determined by a single parameter $\mathfrak e$ describing the valley geometry, independent of doping, scattering time, and other system parameters (Fig.~\ref{fig2}d). This contrasts sharply with conventional VHE, which shows the opposite behavior.
This remarkable feature is due to the fact that both charge and valley responses here arise from the same
mechanism (Drude mechanism here), which naturally leads to their same parametric dependence (on $\mu$ and $\tau$). This remarkable behavior originates from the unique character of TRIVs discussed above, i.e., it is now the $\mathcal T$-even mechanism, rather than the usual $\mathcal T$-odd ones, that contributes to VHE.

From Eq.~(\ref{VHA}), $\theta_{\mathrm{VH}}$ is a monotonically increasing function of $\mathfrak e\in [0,1)$ (see Fig.~\ref{fig2}c). Giant valley Hall angles can be inferred by simply inspecting the valley Fermi surface of a TRIV system. For example, $\mathfrak e\sim 0.8$ can already give a giant $\theta_{\mathrm{VH}}\sim 0.5$. In addition, we note that for weak eccentricity,
Eq.~(\ref{VHA}) can be simplified as
\begin{eqnarray}
\theta_{\mathrm{VH}}\approx \zeta\left|\sin\left(2\phi\right)\right|, \label{VHA-1}
\end{eqnarray}
where $\zeta\equiv \lambda -1>0$ is another parameter characterizing eccentricity.

\begin{table}
\caption{\label{tab1} 2D Bravais lattices and layer groups that are compatible with TRIVs and eccentricity VHE. The second column lists the symmetry $\mathcal{Q}$ that connects the two TRIVs.}

\begin{ruledtabular}
\renewcommand{\arraystretch}{1.5}
\begin{centering}
\begin{tabular}{ccc}
TRIVs & $\mathcal{Q}$ & Generators of little co-group at each TRIV\tabularnewline
\hline
\multirow{5}{*}{\includegraphics[width=1.7cm]{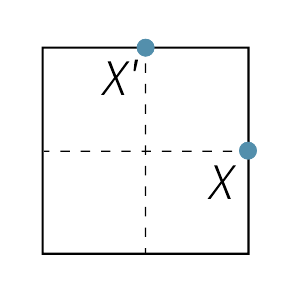}} & $C_{4z}$ & \{$C_{2z}$\} (for LG 49); \{$\mathcal{P}$, $C_{2z}$\} (51,52); \tabularnewline
 &  & \{$C_{2z}$, $C_{2x}$\} (53,54); \{$C_{2z}$, $M_{x}$\} (55,56); \tabularnewline
 &  & \{$\mathcal{P}$, $C_{2z}$, $C_{2x}$\} (61-64)\tabularnewline
 & $S_{4z}$ & \{$C_{2z}$\} (49); \{$C_{2z}$, $C_{2x}$\} (57,58); \tabularnewline
 &  & \{$C_{2z}$, $M_{x}$\} (59,60)\tabularnewline
\cline{2-3}
\multirow{4}{*}{\includegraphics[width=1.7cm]{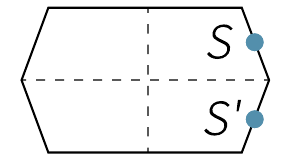}} & $C_{2x}$ & \{$\mathcal{P}$\} (18); \{$C_{2z}$\} (22); \{$\mathcal{P}$, $C_{2z}$\} (47,48)\tabularnewline
 & $M_{x}$ & \{$C_{2z}$\} (26)\tabularnewline
 & $C_{2x}$ & \{$E$\} (10)\tabularnewline
 & $M_{x}$ & \{$E$\} (13); \{$M_{z}$\} (35,36)\tabularnewline
\end{tabular}
\par\end{centering}
\end{ruledtabular}
\end{table}

\paragraph{\textcolor{blue}{Universality in TRIV systems.}}
In deriving Eq.~(\ref{VHA}), we assumed that the two TRIVs are connected by $C_{4z}$ symmetry. Here, we show that
eccentricity VHE and its key features are \emph{universal} for \emph{all} 2D TRIV systems.

First of all, among the five 2D Bravais lattices, only square lattice and centered rectangular lattice can support a pair of symmetry-connected TRIVs.
To see this, one notes that for oblique and rectangular lattices, no symmetry can connect two TRIVs; for hexagonal lattices there are three (rather than two) $C_{3z}$-connected TRIVs (the $M$ points). Therefore, the only possibilities left are the square lattice (TRIVs at $X$ and $X'$) and the centered rectangular lattice (TRIVs at $S$ and $S'$), as shown in Table~\ref{tab1}.

By screening the 25 layer groups for these two Bravais lattices (Supplementary Section I), we find that all of them can support the eccentricity VHE (see Table~\ref{tab1}).
Notably, 21 out of these 25 groups contain $\mathcal{P}$ and/or $C_{2z}$, in which the conventional VHE is strictly forbidden. This shows  eccentricity VHE greatly extends the scope of valley Hall systems.

For square lattices, $X$ and $X'$ valleys are connected by ${C}_{4z}$ or ${S}_{4z}$. If there is additional ${M}_x$ or ${C}_{2x}$ symmetry at $X$, the principal axis of the Fermi ellipse will be fixed along $x$ and $y$, and $\theta_\text{VH}$ is just given by Eqs.~(\ref{VHA},\ref{VHA-1}). If there is no ${M}_x$ nor ${C}_{2x}$ at $X$, the principal axis of Fermi ellipse generally has an angular offset $\phi_{0}$ from crystal axis. In this case, Eqs.~(\ref{VHA},\ref{VHA-1}) are merely modified by replacing $\phi$ with $(\phi+\phi_0)$ (Supplementary Section II). For example, Eq.~(\ref{VHA}) becomes
\begin{equation}
\theta_{\mathrm{VH}}=\frac{\mathfrak e^2}{2-\mathfrak e^2}\left|\sin\left[2(\phi+\phi_{0})\right]\right|.
\end{equation}

For centered rectangular lattices, $S$ and $S'$ valleys are connected by ${M}_x$ or ${C}_{2x}$, and the Fermi ellipse
generally has a nonzero offset angle $\phi_0$.
The eccentricity valley Hall angle can be found as (Supplementary Section II)
\begin{equation}
  \theta_{\mathrm{VH}}=\frac{\mathfrak e^2 \left|\cos(2\phi)\sin(2\phi_0)\right|}{2-2\mathfrak e^2+\mathfrak e^2 [1+
  \cos(2\phi)\sin(2\phi_0)]}.
\end{equation}
In the case of weak eccentricity, it is simplified as
\begin{eqnarray}
\theta_{\mathrm{VH}} & \approx & \zeta\left|\cos\left(2\phi\right)\sin\left(2\phi_0\right)\right|.
\end{eqnarray}

We thus see eccentricity VHE is a universal effect for all TRIV valleytronic platforms.

\begin{figure}[t]
\begin{centering}
\includegraphics[width=8.7cm]{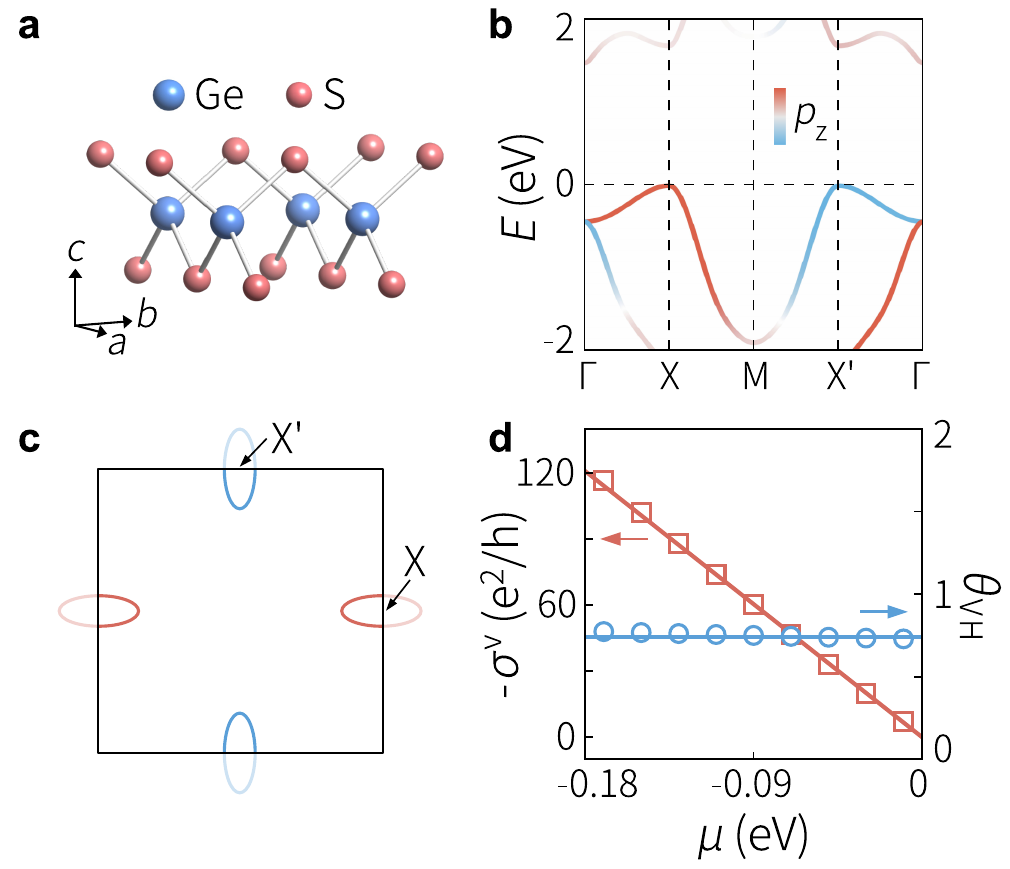}
\par\end{centering}
\caption{\label{fig3}\textbf{Results on monolayer GeS$_2$.} \textbf{a}, Crystal structure of monolayer tetragonal GeS$_2$. \textbf{b}, Calculated band structures. The color map indicates the value of out-of-plane polarization of each state. \textbf{c}, Fermi surface at $\mu =-0.1$~eV. A pair of Fermi ellipses around the two valley centers at $X$ and $X'$ can be seen. \textbf{d}, VHE conductivity and valley Hall angle plotted as functions of chemical potential. The data points are the first-principles results, and the solid lines are results from Eqs.~(\ref{VHC}) and (\ref{VHA}). In calculating $\sigma^v$, we take $\tau=0.1\,$ps.}
\end{figure}

\paragraph{\textcolor{blue}{Material example.}}
To examine the proposed eccentricity VHE, we evaluate the effect in a real material example, monolayer tetragonal GeS$_{2}$,  by first-principles calculations. This material belongs to the group-IV dichalcogenides family. Its bulk form has van der Waals layered structure and has been synthesized in experiments~\citep{Shimada1977Crystallization,Kulikova2014High}. Figure~\ref{fig3}a shows the structure of  monolayer GeS$_{2}$. It has a square lattice, with each Ge at the center of tetrahedra formed by surrounding S atoms. The layer group is 59, meeting the symmetry requirement in Table~\ref{tab1}.

Figure~\ref{fig3}b shows the calculated band structure (see Methods for calculation details). One finds there are a pair of TRIVs at $X$ and $X'$ for the valence band.
The two valley's Fermi ellipses can be clearly seen in Fig.~\ref{fig3}c (plotted at $\mu=-0.1$ eV).
According to Table~\ref{tab1}, the two valleys are connected by $S_{4z}$, and each valley preserves $M_x$. Hence, the eccentricity VHE should be described by Eq.~(\ref{VHA}).

From Fig.~\ref{fig3}b, the two TRIVs can be defined for a wide range of hole doping. Figure~\ref{fig3}d shows the valley Hall conductivity $\sigma^v$ and valley Hall angle as functions of $\mu$ computed from the first-principles band structure, with the driving field along the $[11]$ direction. In comparison, the solid lines are obtained from formulas (\ref{VHC}) and (\ref{VHA}), with eccentricity $\mathfrak e=0.92$. One observes an excellent agreement between them. As expected, $\theta_\text{VH}$ is robust against change in $\mu$, and its robustness against temperature variation (up to room temperature) is shown in
Supplementary Fig.~S1. The obtained $\theta_\text{VH}\sim 0.74$ is also a giant value.

\paragraph{\textcolor{blue}{Discussion.}}
We have proposed TRIVs as a new arena for valleytronics, featuring novel phenomena and innovative approaches to manipulating valley degree of freedom. A fascinating effect unveiled here is eccentricity VHE, which is determined by a geometric parameter $\mathfrak e$ of Fermi surface and is universal across all TRIV systems. Its distinctive features, including the robust geometric valley Hall angle independent of temperature and carrier density, manifest the distinct fundamental mechanism involved in valley transport. Notably, for conventional VHE from Berry-phase mechanism, the response reflects interband coherence hence strongly depends on the band gap, with $\theta_\text{VH}$ quickly suppressed with increasing gap size~\cite{Xiao2007Valley}. In contrast, eccentricity VHE depends only on Fermi surface geometry and not the gap, so it can achieve giant
$\theta_\text{VH}$  even for mid- and large-gap semiconductors, as evidenced by the example GeS$_2$. This is a big advantage for applications.

Interestingly, TRIV systems include the layer groups that also support valley-layer coupling~\citep{Yu2020Valley} (including layer group 10, 22, 50, 59, and 60). In such cases, states of two TRIVs have nonzero and opposite out-of-plane charge polarizations, thus valley polarization can be controlled by a gate field~\citep{Yu2020Valley}. Monolayer GeS$_2$ is such an example, with its out-of-plane polarization (Methods) plotted in Fig.~\ref{fig3}b. This offers an additional route for valley control.

The eccentricity VHE induces valley polarization accumulated around boundaries normal to the valley current flow direction, which can be experimentally detected via optical linear dichroism. Alternatively, valley polarization can be generated first, e.g., using linearly polarized light or a gate field in systems with valley-layer coupling, leading to a transverse charge current whose sign reflects the valley polarization and which is readily measurable.

\begin{figure}
\begin{centering}
\includegraphics[width=8.7cm]{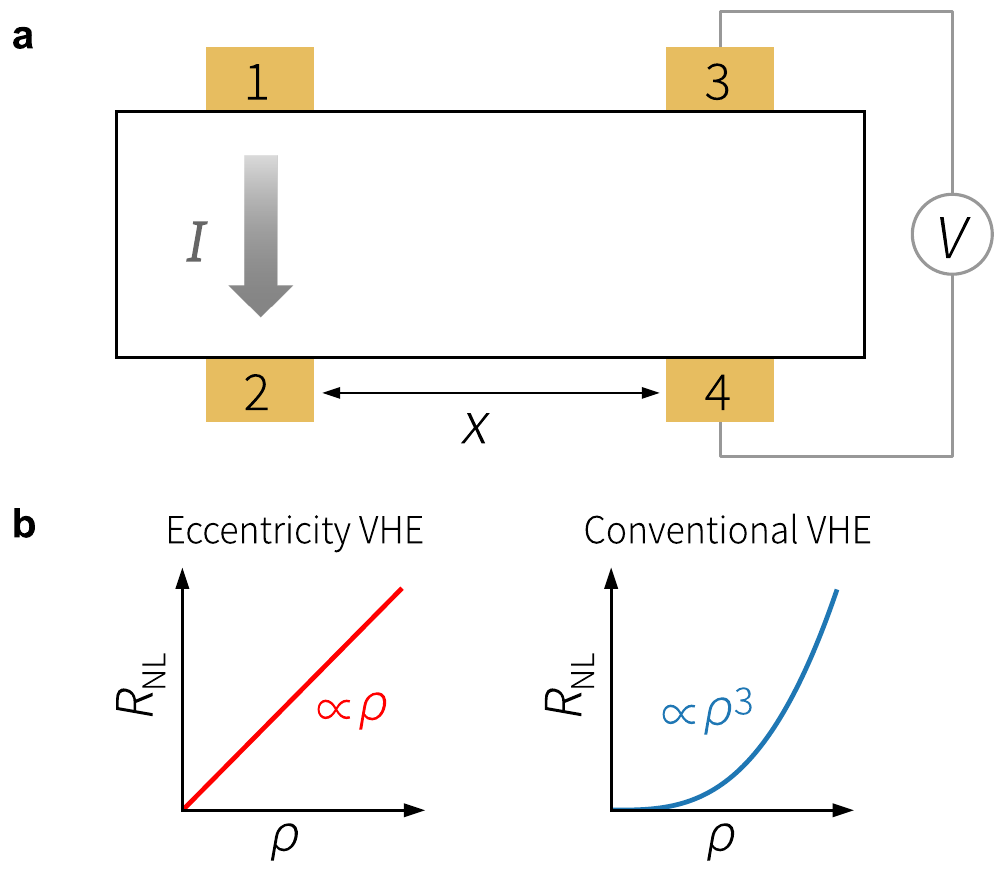}
\par\end{centering}
\caption{\label{fig4}\textbf{Nonlocal transport signature.} \textbf{a}, Experimental setup for nonlocal measurement of VHE.
A charge current is applied between contacts 1 and 2, and the resulting voltage is measured between contacts 3 and 4 at a distance $x$ along the sample strip.
\textbf{b}, Eccentricity VHE features a nonlocal resistance $R_\text{NL}\propto \rho$, in contrast to the $R_\text{NL}\propto \rho^3$ scaling for conventional VHE. }
\end{figure}

Another more direct method to detect eccentricity VHE is via nonlocal measurement, with the typical setup illustrated in Fig.~\ref{fig4}a. A driving current $I$ is applied between electrodes 1 and 2, and at a distance of $x$ away, a nonlocal voltage $V_\text{NL}$ is detected between 3 and 4. For eccentricity VHE, one finds the following relation for the nonlocal resistance (Supplementary Section IV)
\begin{equation}
R_\text{NL}=V_{\mathrm{NL}}/I\sim \rho \theta_{\mathrm{VH}}^2 e^{-x/\ell_{v}},
\label{nonlocal}
\end{equation}
where $\rho$ is the charge resistivity and $\ell_{v}$ is the valley diffusion length.
This result is in sharp contrast to the conventional VHE which has a scaling $R_\text{NL}\propto \rho^3$~\cite{Gorbachev2014Detecting,Sui2015Gate,Shimazaki2015Generation}. By fitting experimental data, this also offers a direct way to evaluate the eccentricity valley Hall angle $\theta_{\mathrm{VH}}$.

Finally, beyond VHE, TRIV systems  may host a broad spectrum of intriguing valley phenomena --- for example, valley transport and polarization driven by thermoelectric or nonlinear mechanisms. The expanded material classes, enriched valley–layer–spin couplings, and distinct symmetry constraints together suggest  a new arena for valleytronics research.

\section{Methods}
\textcolor{blue}{First-principles calculations.}
The band structure of GeS$_2$ was calculated using density functional theory as implemented in the VASP package~\cite{Kresse1993Ab,Kresse1996Efficiency,Kresse1996Efficient}. The projector augmented-wave method was employed, with a plane-wave energy cutoff of 400~eV~\cite{Bloechl1994Projector}. The Perdew–Burke–Ernzerhof treatment~\cite{Perdew1996Generalized} was used to model the exchange-correction functional. The lattice constant was set to $a=3.456$~{\AA}~\cite{Shimada1977Crystallization}. A vacuum layer of thickness of 15~{\AA} was applied to suppress the artificial interaction from periodic images. The energy convergence criterion was set as 10$^{-6}$~eV. For Brillouin zone sampling, a $\Gamma$-centered $k$-point mesh with size of $16\times16\times1$ was used. The \textit{ab initio} tight-binding models were constructed using the Wannier90 package~\cite{Mostofi2014updated}. The $s$ and $p$ orbitals of Ge atoms, and the $s$ and $p$ orbitals of S atoms were used as the initial input for the local basis.

\textcolor{blue}{Layer polarization.}
In Fig.~\ref{fig3}b, the out-of-plane polarization for a Bloch state $|u_{n\boldsymbol{k}}\rangle$ is evaluated as
$
p_{z}=\left\langle u_{n\boldsymbol{k}}|\hat{z}|u_{n\boldsymbol{k}}\right\rangle.
$
Here, the origin $z=0$ is set at the middle point of the thickness, i.e., at the Ge atomic layer for the monolayer GeS$_2$ structure. Then, a state with $p_{z}>0$ $(<0)$ indicates that the state has more distribution in the upper (lower) atomic layer. The little co-group of the $X$ ($X'$) valleys is generated by $C_{2z}$ and $M_x$ (see Table~\ref{tab1}), which allows a finite $p_z$ at the two valleys. Meanwhile, the two valleys are connected by the $\mathcal{Q}=S_{4z}$ symmetry, which requires the two valleys have opposite $p_z$, as shown in Fig.~\ref{fig3}b. Based on this feature, a gate electric field can induce a valley splitting and generate a valley carrier polarization.

%

\end{document}